\documentclass[%
 aip,
 sd,%
 amsmath,amssymb,
 reprint,%
]{revtex4-1}

\usepackage{graphicx}
\usepackage{dcolumn}
\usepackage{bm}

\begin{document}

\preprint{AIP/123-QED}

\title[Local minimum in effective pairpotentials]{Local minimum in effective pairpotentials: Pseudopotential theory revisited}
\author{G. M. Bhuiyan}
\email{gbhuiyan@du.ac.bd} 
\affiliation{Department of Theoretical Physics, University of Dhaka, Bangladesh. }
\author{Fysol Ibna Abbas}
\affiliation{Department of Theoretical Physics, University of Dhaka, Bangladesh.}%

\date{\today}
\begin{abstract}
Local minimum appearing in the interionic pair interactions, when derived from local model pseudopotential, for
Al (and some other polyvalent metals) remains as a long standing problem of clear understanding 
although some attempts are made by different authors. The origin of this feature of local minimum is systematically
investigated in this article, considering both the chemical valence, Z, and the core radius, $R_{c}$ as variables.
Ashcroft's empty core model is used to describe the interionic pair-potential, because, it depends on these two parameters only. Results of this investigation show that monovalent metals do not exhibit a local minimum at small $r$ but some polyvalent metals does where, the core radius plays the major role.
\end{abstract}
\pacs{60; 70}
\maketitle
\section{Introduction}
According to the pseudopotential theory, effectively started in sixties of the last century, the 
electron-ion pair pseudopotentials are treated in two ways [1]. One is the ab-initio type, derived employing  the first principles right from the atomic level description[2,3] and the second types are the model pseudopotentials proposed by many authors [4,5,6,7,8] which are generally parameter dependent. Among the model pseudopotentials the Ashcroft's empty core model [4] and its kins [7] are widely used, and hundreds of articles studying physical properties of simple metals and  their alloys have been published so far. The advantage of the empty core model is that it has a single paratermeter, known as core radius, $R_{c}$, to be fixed for effective calculations. In some cases, specially for transition metals, the effective valance, $Z$, acts as parameter to account for the effects of sd-mixing [6,9,10]. However, these parameters are suitably adjusted either by fitting to experimental data for different  physical properties [11,12,13] or by reproducing desired magnitude of physical properties theoretically [7,9,10]. 

Hafner and coworkers [14,15,16] studied structural and electronic properties of many polyvalent metals using the empty 
core [4] and the optimized pseudopotentials. They derived interionic interactions from these pseudopotentials with Vasishta and Singwi [17], or Ichimaru and Utsumi [18] dielectric functions including even relativistic or non-relativistic core orbitals. In all cases they found a positive local minima at small $r$ near the first nearest neighbour distance in solid [15] and  in liquid phase for Al. Similar trends were also found for some other polyvalent metals such as In, Zn, etc. According to Jank and Hafner [15], the strong electron-ion interaction leads to high electron density and as a result, the first attractive minimum closed to nearest neighbour distance becomes quite shallow and, the minimum is partically coverd by the repulsive core. Now question remains, if the repulsive core be much stronger would the attractive minimum go up to yield a repulsive local minimum as in the clear case of Al? This question is not addressed in [15] but a weak explanation of shifting the position of the principal minimum is given by Hafner and Jank that "as the electron density increases and/or the core radius decreases, the minimum is shifted relative to the nearest neighbour distance and gets flattend". Hafner and Jank also suggested that, the essential trend may be parameterized  in terms of $Rc$/$Rs$ ($Rs$ being the electron density parameter which is sometimes refered to as Wegner-Seitz radius) ratio.

It is surprizing why no one, until now, attempted to explain the origin of the charecteristic feature of repulsive  local minimum specifically for Al during the last decades even after encountering the local minimum theoretically for the first time. We in the present article, intend to address the root cause of this characteristic feature  found in effective pair potentials. Specifically, we shall investigate the role of the core size and the chemical valence, and how do they affect the minimum of the effective interparticle pair intarections. Our samples  to study starts from monovalent metals and ends to pentavalent ones, namely Na, K, Zn, Al, Sn and Bi, respectively. 

The layout of this paper is the  following. The empty core pseudopotential and relevant screening functions are described in section 2. Section 3 is devoted for the results and discussion. We conclude this article in section 4.
\section{Theory}
The main ingredients of the theory are a model potential and the dielectric function obtained from the linear response theory.
\subsection {The empty core model}
The empty core model proposed by Ashcroft reads(in atomic unit)[4]
\begin{equation}
W_{b}(r) =\begin{cases}
0 & \text{if $r<R_{c}$,} \\
-\frac{Z}{r} & \text{if $r>R_{c}$},
\end{cases}
\end{equation}
where $W_{b}(r)$ is the electron-ion interaction, $R_{c}$ the core radius, $Z$ the s-electron effective valance, and $e$ the electronic charge. The unscreened form factor of (1) is[19,20] 
\begin{eqnarray}
v(q)=- \frac{4 \pi Z n}{q^{2}} \cos(qR_{c})
\end{eqnarray}
where $n$ is the ionic number density, and $q$ the momentum transfer. Please note that the core radius, $R_{c}$,
enters into the form factor of the interaction through the scattering matrix. This form factor finally carries the
core radius into the interionic interaction (see below).

Within the pseudopotential formalism, the effective interionic pair interaction can be 
expresed as [19]
\begin{eqnarray}
v_{eff}(r)= \frac{Z^{2}}{r} \left[1-\frac{2}{\pi} \int I^{J} \,\frac{\sin(qr)}{q} dq\right]
\end{eqnarray}
here, the energy wave number chracteristic
\begin{eqnarray}
I^{J}=\left[\frac{q^{2}}{\pi n Z}\right]^{2} |V(q)|^{2} \left[1-\frac{1} {\epsilon(q)}\right] \left[1-\frac{1}{1-G(q)}\right],
\end{eqnarray}
where $\epsilon(q)$ is the dielectric function for the electrons and $G(q)$ the local field correction. These dielectric functions are taken from the Ichimaro and Utsumi[18] because their functions satisfy the compressibility sum rule. In equation (3) the first and the second term on the right hand side represent the direct and indirect interactions respectively. The direct term is just the Coulomb repulsion between two ions and the indirect term provides the attractive interaction between ions which is mediated by conduction electron density. 

\section{Result and Discussion} 
We are interested to investigate the origin of the creation of local minimum in the effective pair potentials of polyvalent metals, in particular, in Al, at small $r$. In order to examine this feature systematically we have started calculations for the monovalent metals such as Na and K. Then gradually advances toward the divalent (Zn), trivalent (Al), tetravalent (Sn) and pentavalent (Bi) metals.
\begin{figure}[h!]
(a)\includegraphics[width=5cm,height=5cm, angle=270]{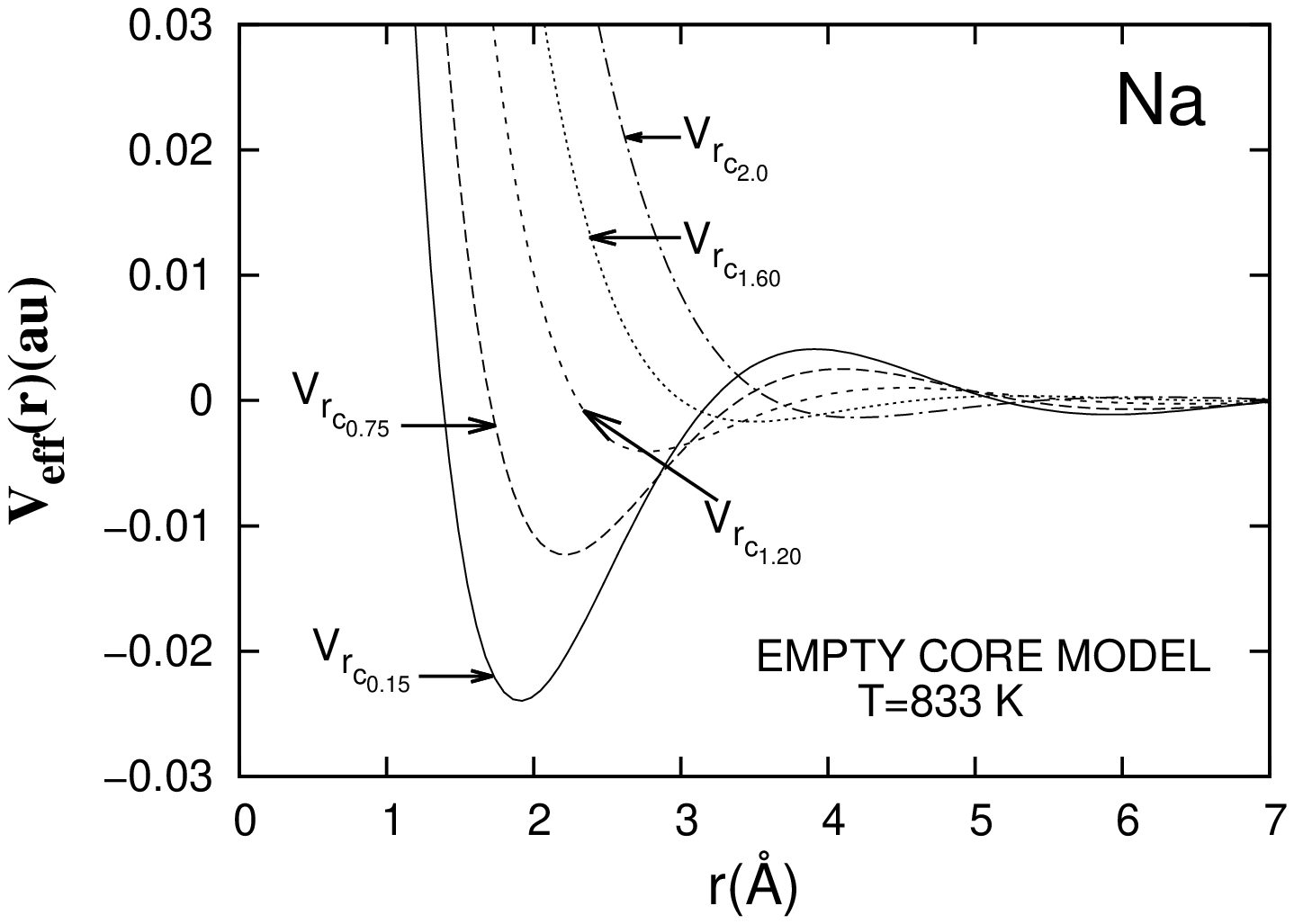}
(b)\includegraphics[width=5cm,height=5cm,angle=270]{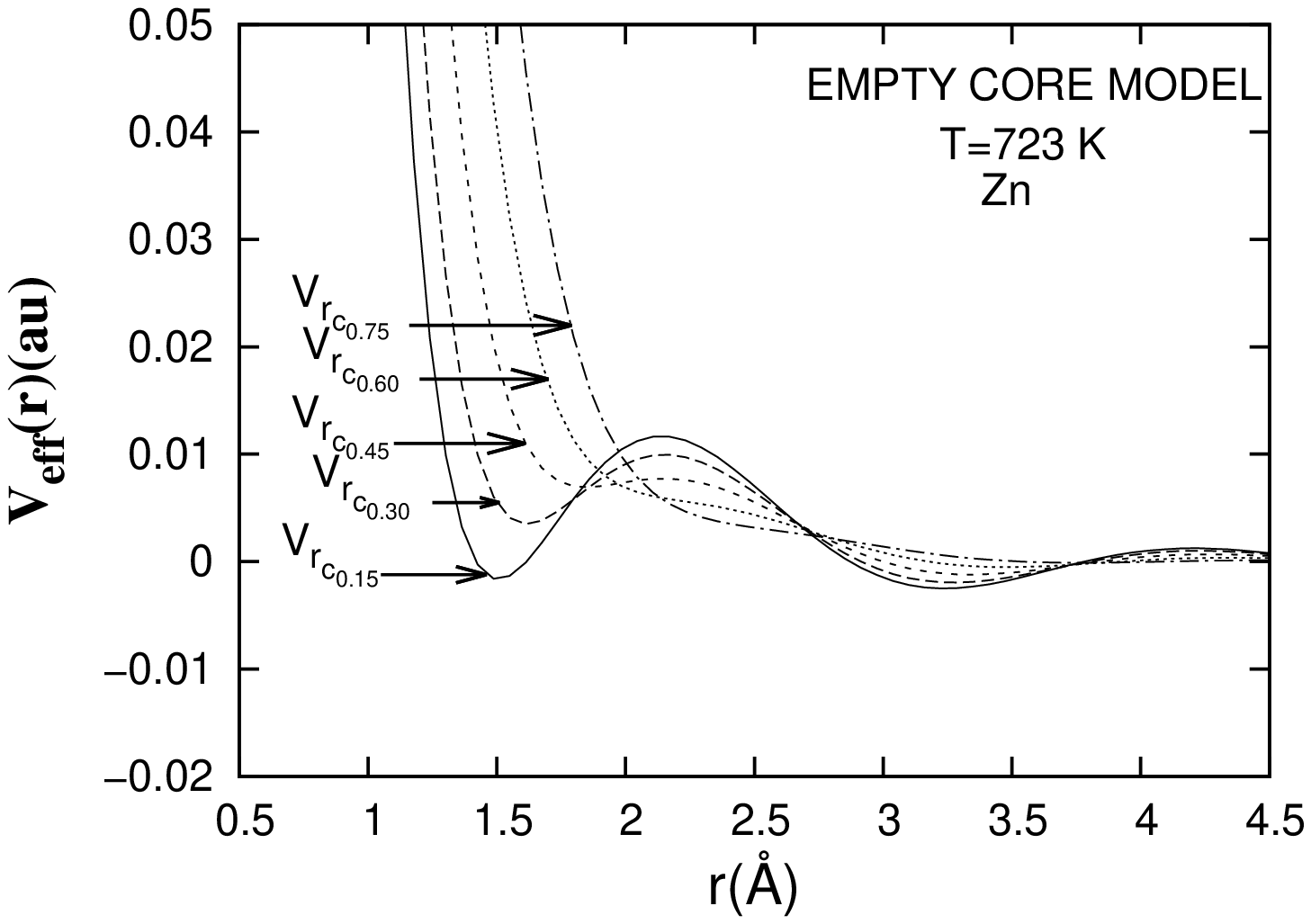}
\caption{Effective pair potentials, $v_{eff}(r)$, as a function of $r$ for different $R_{c}$ values; (a) Na (b) Zn.}
\end{figure}

Figures 1(a) and 1(b) show the effective pair potentials, $v_{eff}(r)$, for monovalent Na and divalent Zn as a function of  $r$ for different values of $R_{c}$. It is noticed that the potential well for Na is the deepest for the smallest value of core size that is for the  smallest value of core radius, $R_{c}$ (=0.015 $\AA$). As core radius increases the depth of the well reduces gradually and at the same time the position of the minimum shifts toward large $r$. But the value of the minimum remains always negetive and no local minimum appears. Hafner and Jank [15] argued  that as the core radius decreases the electron density increases and the parameter defined as ($R_{c}/R_{s}$) ($R_{S}$ being the Weigner-Sitz radius) can explain the essential trend of the pseudopotential. Please note that, the average conduction electron density, in this study, remains constant as the volume of the sample is kept same regardless of the core size is changed or not. We have also observed similar feature in $v_{eff}(r)$ for monovalent metal K. 
\begin{figure}[h!]
(a)\includegraphics[width=5cm,height=5cm,angle=270]{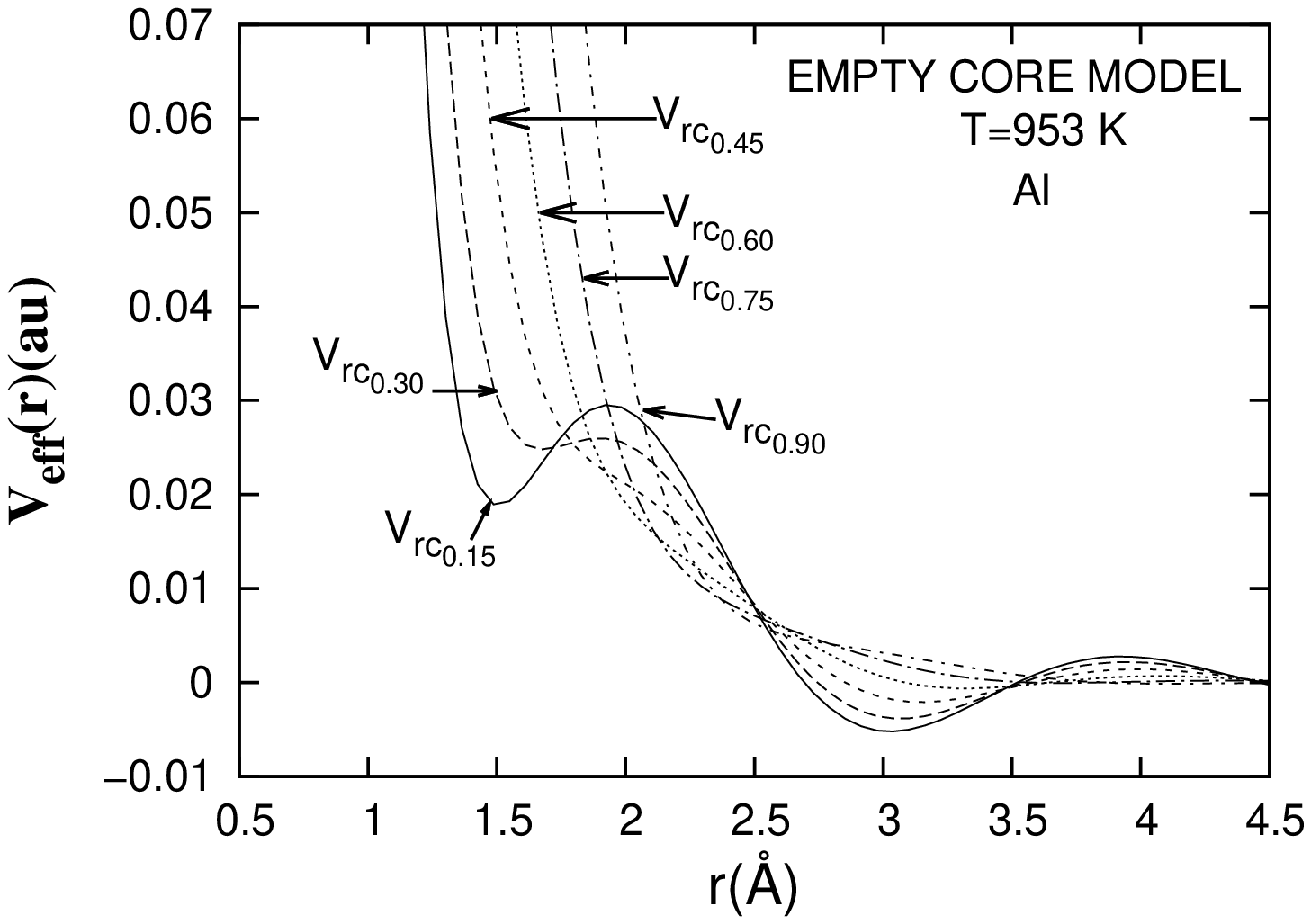}
(b)\includegraphics[width=5cm,height=5cm,angle=270]{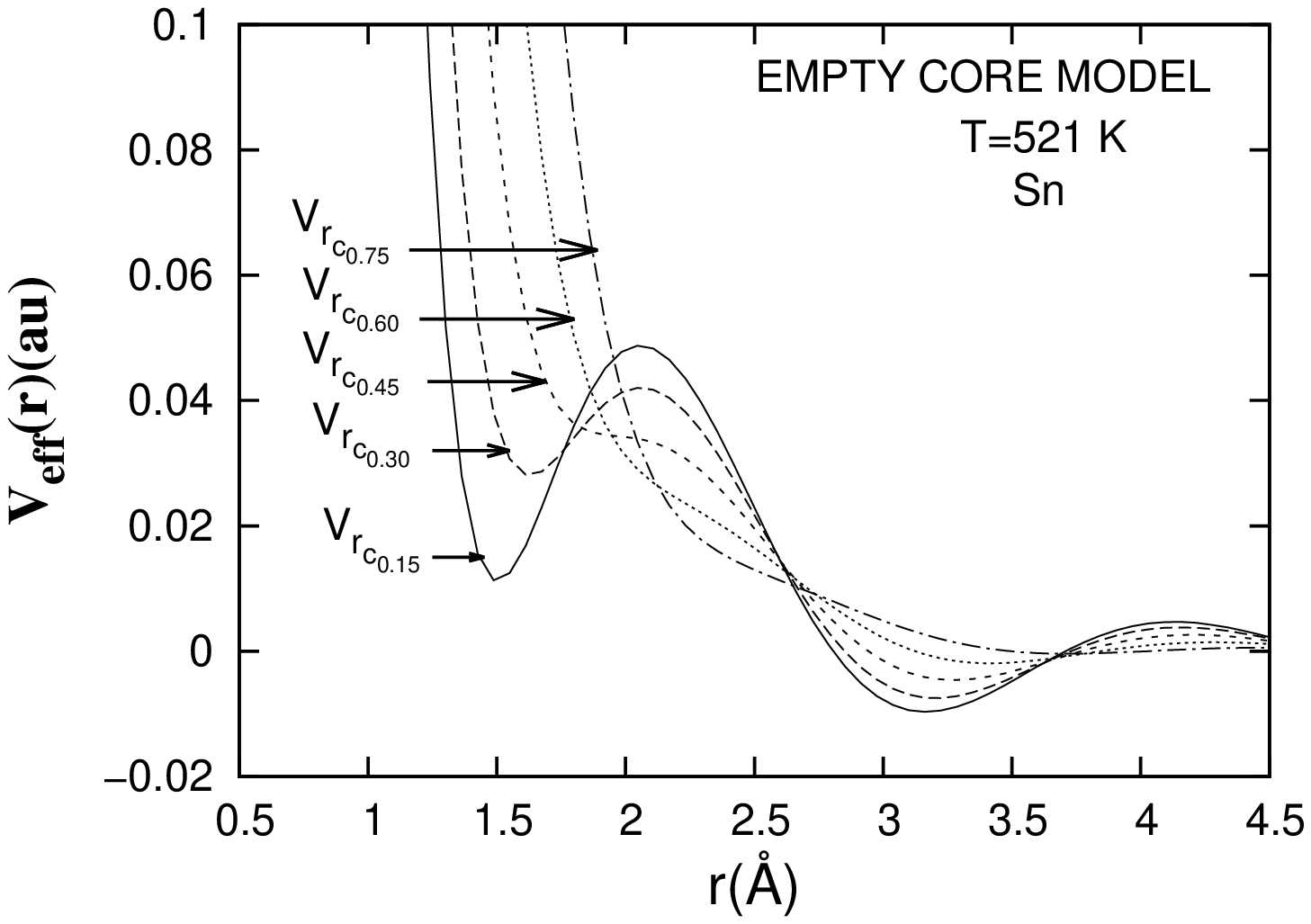}
(c)\includegraphics[width=5cm,height=5cm,angle=270]{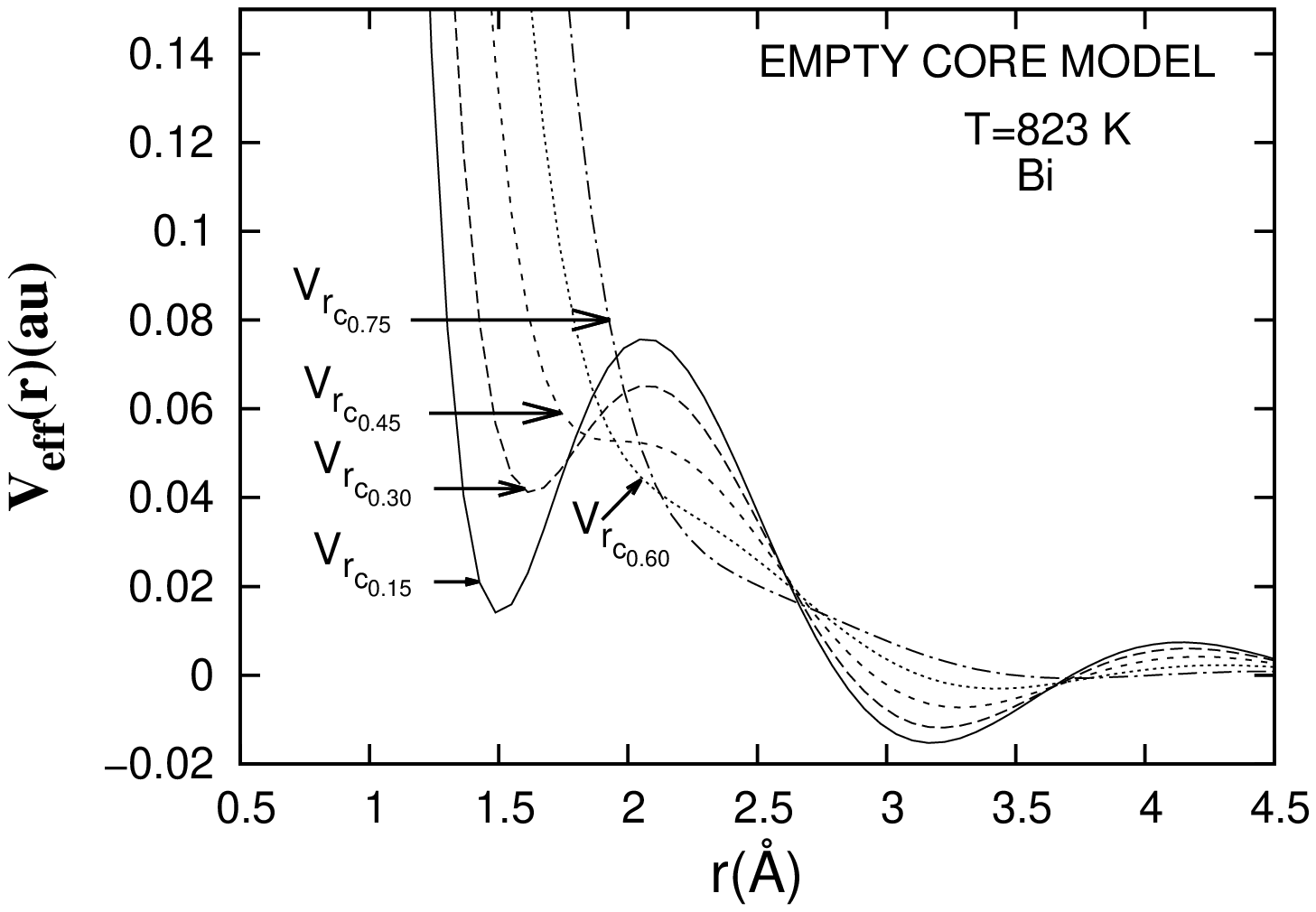}
\caption{Effective pair potentials, $v_{eff}(r)$, as a function of $r$ for different $R_{c}$ values; (a) for Al (b) for Sn and (c) for Bi.}
\end{figure}
Figure 1(b) shows how the depth of the well reduces with increasing value of $R_{c}$ for diatomic metal Zn.
\begin{figure}[h!]
\includegraphics[width=5cm,height=5cm]{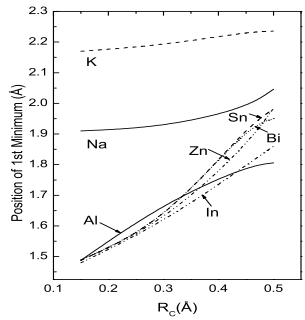}
\caption{Position of minimum of the first well of $v_{eff}(r)$ as a function of $R_{c}$.}
\end{figure}
It is noticed that the depth of potential well reduces and the position of the well shifts  with increasing $R_{c}$ as in the case of monoatomic metals. But in this case a difference is noticed; the first minimum crosses the zero level and goes up to the positive region to yield a local minimum. Farther increase of $R_{c}$ value cause the local minimum to be disappeared. Consequently, the second minimum turns to be as the principal minimum of the effective potential.

\begin{figure}[h!]
\includegraphics[width=6cm,height=6cm]{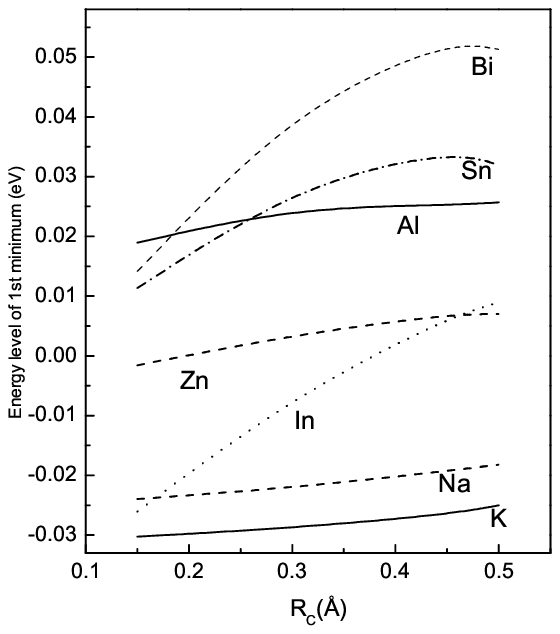}
\caption{Energy level of the minimum of the first well of $v_{eff}(r)$ as a function of $R_{c}$}
\end{figure}
Figure 2(a) shows $V_{eff}$ as a function of $r$ for the trivalent metal Al. Here it is seen that the first minimum lies in the positive region of V(r) for the smallest core size and it behaves as like a local minimum. The energy level 
of the minimum goes up and and at the same time shifts to large $r$ as $R_{c}$ values increases. After certain value of 
$R_{c}$, which may be referred to as critical value $R_{c}^{c}$,  this local minimum disappears from the scene completely, and the second minimum then becomes the principal minimum. The underlying cause of this disappearness of the local minimum for $R_{c}> R_{c}^{c}$ is discussed later. It is worth noting here that the depth of the second minimum also reduces and shifts toward large $r$ with further increase of $R_{c}$, but the well does not cross the zero energy level. Similar trends
of chnging in the local minimum are also found in the case of tetravalent (Sn)  and pentavalent (Bi) metals (see Figures 2(b) and 2(c)). But in the latter cases the positive energy level of the well of the local minimum goes to higher and higher positions. Other minor differences, in the behaviour of local minimum, between different polyvalent metals are also discussed in figures 3 and 4.

Let us now look at what are the underlying mechanisms involved in forming the local minimum in the effective pairpotentials of some polyvalent metals. From the global perspective of the metallic sample, increment of the core size reduces the 
interionic space for conduction electrons and this in turn increases conduction electron density. According to the density functional theory kinetic energy of electrons increases with increasing density due to Pauli's repulsion. So, with increasing core size kinetic enery of conduction electrons increases. Consequently, the attractive indirect interaction term of the pseudopotential, which is mostly responsible for producing potential well, increases through the scattering matrix involved therein. By the term increase of interaction we mean that it becomes less attractive. Obviously, the depth of the well of $v_{eff}(r)$ reduces due to interplay between the repulsive direct and attractive indirect interactions. The interesting thing is that, as the core size increases the principal well goes up gradually 
keeping the well in the memory and it continues until the critical value of $R_{c}^{c}$ is reached. For $R_{c}> R_{c}^{c}$
kinetic enery of conduction electrons becomes so large that the local minimum that remains in the memory is totally lost.

In forming the local minimum, the effective valence, $Z$, also plays a significant role as well. Because, polyvalent metals
($Z>1$) gives off more electrons in the conduction band than monovalent ones. Thus it affects the kinetic energy of the system through the density of the conduction electrons. As a result monovalent metals (i,e. $Z=1$) does not show a local minimum in $v_{eff}(r)$ whatever be the size of the core, whereas some polyvalent metals does.  

Figure 3 shows the shifting of the first minimum of the $v_{eff}(r)$ for different values of $R_{c}$ and for different metals. 
In a general view it is seen that the slope of shifting w.r.t. $R_{c}$ is the smallest for monoatomic metals Na and K
and, it is much larger for polyvalent metals. That is, increase of $R_{c}$ value cause polyvalent metals to shift
position of minimum of $v_{eff}(r)$ to large $r$ much strongly than monovalent ones. 

Figure 4 illustrates how the energy level of the principal minimum varies with the change of $R_{c}$ values. In this case it appears that Na, K, Zn, and Al show similar behaviour in this regard. On the other hand, In, Sn and Bi show the same pattern with relatively larger change of energy level with variation of $R_{c}$ than the former ones. In the former group metals are monovalent, divalent anf trivalent, and in the later group metals are trivalent (with larger atomic mass), tetravalent and pentavalent.
From figure it is also noticed that, for the same smallest value of $R_{c}(=0.015\, \AA)$ energy level of the first minimum is positive for Al, Sn and Bi, whereas it is negative for the others. That is, polyvalent metals Al, Sn and Bi exhibit
 local minima even starting from the smallest core size chosen and the others show at larger values of $R_{c}$.
\section{Conclusion} We have systimatically investigated  for the first time the effect of the core size 
parameter of the ion on the  interionic pair interaction of monovalent and polyvalent metals. The local pseudopotential is described by the Ashcraft epmty core model[4] and the dielectric function by the Ichimaro-Utsumi theory [17]. From the systematic analysis of the role of core radius $R_{c}$ we can draw the following conclusions.
(i) For monovalent metals (Na and K) the depth of the first minimum of $v_{eff}(r)$ reduces and at the same time position of the minimum shifts toward large r with increasing values of $R_{c}$. But any local minimum at the positive energy level does not appear at all for monovalent metals.
(ii) For polyvalent metals initialy the trends are similar to that of the monovalent metals, i.e. the depth reduces and the position of the first minimum shifts with increasing $R_{c}$. But when the value of $R_{c}$  becomes greater than a certain magnitude, the minimum of the first well moves to a positive energy level and behaves like a local minimum at small $r$. The second minimum of $v_{eff}(r)$ is then behaves like the principal minimum and is followed by Friedel oscillations. However further increment  of $R_{c}$ cause the local minimum to disappear  complelely.
(iii) For the same core radius, the energetic position of the local minimum in polyvalent metals goes to higher and  higher levels with increasing valence, $Z$.
(iv) The characteristic feature of local minimum in the interionic pseudopotential for Al and other polyvalent metals 
largely depends on the value of core radius $R_{c}$, the chemical valence also plays a
role in this process. \\ \\         

References\\ \\
$[1]$ W A Harrison, Pseudopotential in the Theory of Metals (W. A. Benjamin, 1966)\\
$[2]$ R Car and M Parninello. Phys Rev Lett. 55(1966) 2471\\
$[3]$ Martin Fuchs and Matthias Scheffler, Comp. Phys. Comm. 119 (1999)67.\\
$[4]$ N W Ashcroft, Phys. Lett. 23(1966)48.\\
$[5]$ V Heine and Averenkov, Phil. Mag. 9(1964)451\\
$[6]$ J M Wills and W.A Harrison, Phys  Aev. B28(1983)4363.\\
$[7]$ J L Bretonnet and M Silbert, Phys Chem. Liq. 24(1992)169.\\
$[8]$ W H Young, Rev. Prog. Phys, 55(1992) 1769.\\
$[9]$ G M Bhuiyan, J. L. Bretonnet, L. E. Gonzalez and M. Silbert, J. Phys.: Condens. Matter 4 (1992) 7651.  \\
$[10]$ G M Bhuiyan, J.L Bretonnet and M. Silbert, J. Non-cryst. Solid 156-158(1993)145.\\
$[11]$ C Regnaut, Z. Phys. B76, (1989) 179.\\
$[12]$ Ch Hausleitner, G. Kahl and J. Hafner, J. Phy: Condens. Matter 3(1991) 1589.\\
$[13]$ N W Ashcroft and D C Langreth, Phys. Rev. 156 (1967) 685.\\
$[14]$ J Hafner and W Jank, Phys Rev. B42(1990-II)11530.\\
$[15]$ W Jank and J. Hafner, Phy Rev. B 42(1990-I) 6926.\\
$[16]$ W Jank and J. Hafner, Phy. Rev B41, (1990-II)1497.\\
$[17]$ P Vashishta andn K.S. S Singwi, Phys. Dev. B6 (1972) 875.\\
$[18]$ S Ichimaru and K. Utsumi, Phys. Rev. b24(1981) 7385.\\
$[19]$ M Shimoji, Liquid Matals (Academic Press,1976)p 96.\\
$[20]$ J Hafner, From Hamiltonians to Phase Diagrams, (Springer- Verleg, Berlin 1987).

\end{document}